\documentclass[11pt, oneside]{article}   	% use "amsart" instead of "article" for AMSLaTeX format
\usepackage{geometry}                		% See geometry.pdf to learn the layout options. There are lots.
\usepackage{graphicx}				% Use pdf, png, jpg, or eps§ with pdflatex; use eps in DVI mode
\usepackage{amssymb}
\usepackage{hyperref}
\usepackage{natbib}

     \usepackage[full]{textcomp}
     \usepackage[osf,sups]{Baskervaldx} % osf for text, not math
     \usepackage{cabin} % sans serif
     \usepackage[varqu,varl]{inconsolata} % sans serif typewriter
     \usepackage[baskervaldx,bigdelims,vvarbb]{newtxmath} % bb from STIX
     \usepackage[cal=boondoxo]{mathalfa} % mathcal

\DeclareRobustCommand{\ion}[2]{\textup{#1\,\textsc{\lowercase{#2}}}}

\title{Alexander Dalgarno and the development of astrochemistry}
\author{John H.~Black \\
Professor Emeritus \\
Chalmers University of Technology, Sweden \\
{\it john.black@chalmers.se}  \\
this article was prepared to accompany a talk at a \\
meeting on the History of Astrochemistry organized by the \\ 
Historical Group of the Royal Society of Chemistry \\
Burlington House, London, 2025 October 16 
}
\date{}							% Activate to display a given date or no date

\begin{document}
\maketitle
\section{Introduction}
The interdisciplinary field of astrochemistry arose during the 1970s as observations in previously unexplored parts of the electromagnetic spectrum began to reveal the extent of a molecular component of interstellar matter with a surprisingly rich chemistry. Astrochemistry expanded further in order to explain the role of atomic and molecular processes in a broad range of phenomena in the universe. It is instructive to describe the current scope of astrochemistry using the career and accomplishments of Alexander Dalgarno as an organizing principle. Dalgarno helped to establish a self-sustaining community of astrochemists around the world. His own research interests highlight the early development of astrochemistry and anticipate much of its later evolution. His theoretical investigations of fundamental atomic and molecular processes lie at the heart of the subject. 

\section{Background}
The life and career of Alex Dalgarno have been summarized in many articles. Only a few are mentioned here. 
The Royal Society published a comprehensive biographical memoir of Alexander Dalgarno \citep{2020BMFRS..69..145H}, the online version of which includes a link to an authoritative list of publications. Also worth noting are the autobiographical article \citep{2008ARA&A..46....1D} and the transcript of an interview that is part of the oral history program of the American Institute of Physics.\footnote{Interview by David DeVorkin on 2007 December 6, Niels Bohr Library \&\ Archives, American Institute of Physics, College Park, MD USA \href{https://repository.aip.org/node/128481}{{\tt https://repository.aip.org/node/128481}}.}

Briefly, Alex Dalgarno was born in London in 1928. He studied at University College London and completed his PhD in 1951, after which he joined the faculty of the Queen's University of Belfast and there helped David Bates to establish a research group in theoretical atomic physics. At Belfast and through visits to the USA, he was exposed to research on atomic and molecular processes in the upper atmosphere of the Earth and to electronic computers. In the early 1960s he began to apply his knowledge of atomic physics to some astrophysical problems. In 1967 he was appointed Professor of Astronomy at Harvard University and Staff Scientist at the Smithsonian Astrophysical Observatory in Cambridge MA, USA. His long and extraordinarily productive career continued there until his passing in 2015. 

\section{Advances in observation and technology}
Alex Dalgarno was a theoretician. He recalled a single experience of observation when his PhD student Paul Kalaghan persuaded him to follow along to the Haystack 37 meter radio telescope for an observing session.
% for an unsuccessful search for hyperfine structure lines of vibrationally excited H$_2^+&. 
Nonetheless he often emphasized that astronomy is an observational science and he followed very closely the latest observations relating to molecules in space. No doubt his position as Letters Editor of {\it The Astrophysical Journal} (1973-2002) \citep{2003ApJ...582L...1D} offered a panoramic view of important developments as they happened. It also put him in contact with referees and their critical assessments of progress in astrochemistry. The anecdotal record also indicates that he had to intervene as `referee' himself in some pugilistic disputes over priority in discovery and publication. Despite his sense of humor, his editorial obligations required him to remove Jack Daniel{\texttrademark} as a co-author of the discovery paper about interstellar ethanol \citep{1975ApJ...196L..99Z}. 

To appreciate fully Dalgarno's contributions it is useful to outline the background of advances in technology and astronomical observation that provided so much of the subject matter in the study of molecules in space. 
\subsection{State of the art in the 1960s}
Prior to 1963, observations in astronomy were largely restricted to the narrow waveband of visible light and a broad swath of the radio spectrum. The interstellar medium was known to contain three molecules, CH, CH$^+$, and CN, based upon the narrow absorption lines that they superimposed on the blue-violet spectra of background stars. Traces of atomic ions (sodium, potassium, calcium) accompanied these species. Solid dust particles in interstellar space were known to cause a general wavelength-dependent extinction of starlight. The spectrum line at 21 cm wavelength (1420.405 MHz frequency), caused by the magnetic-dipole transition between the hyperfine sublevels of the ground state of atomic hydrogen, gave radio astronomers a powerful tool to survey the atomic components of the interstellar medium \citep{1951Natur.168..356E}\footnote{Edward Purcell (1912-1997), who shared the Nobel Prize in Physics for nuclear magnetic resonance and was a Harvard colleague of Dalgarno, returned occasionally to astrophysical problems, as we will see later in the sections on dust and laboratory astrophysics.}.
Dalgarno was the first to perform an accurate quantum mechanical calculation of the cross sections for collision-induced spin change, which are necessary for the interpretation of the \ion{H}{i} 21 cm line\citep{1961RSPSA.262..132D,1969ApJ...158..423A}.
Van de Hulst's \citep{1945NTvN...11..210V} prediction of the 21 cm line also included a discussion of the hydrogenic Rydberg transitions between adjacent levels of high principal quantum number at radio frequencies. Commonly called radio recombination lines, these signatures of highly excited atoms were discovered in the ionized component of the interstellar medium in the mid-1960s. 

Microwave spectroscopy with radio telescopes revealed the presence of interstellar OH (1963), NH$_3$ (1968), H$_2$O (1969), and H$_2$CO (1969). Carl Heiles \citep{1968ApJ...151..919H,1969ApJ...157..123H,1971ARA&A...9..293H} and others began to sense the extent of nearby molecular clouds through OH emission and to infer that these were quite massive gravitationally bound structures. Both OH and H$_2$O were found to appear as natural masers in some circumstances.

\subsection{1970s}
Improvements in the high-frequency performance of radio telescopes and in the sensitivity of mm-wave heterodyne receivers led to an explosion of molecular discoveries. It was now possible to observe strong rotational transitions of abundant molecules like CO as well as uncommon ions and radicals like HCO$^+$ and C$_2$H. Because CO was abundant and easily excited at interstellar densities and temperatures, it became possible to map the extent and distribution of molecular gas throughout the Milky Way (the Galaxy) by means of the emission lines in the pure rotational spectrum of this molecule. Previously unrecognized giant molecular clouds were identified, structures with masses up to one million times the mass of the sun ($10^6$ M$_{\odot}$). Their connections to regions of active star formation were demonstrated. 

Telescopes in orbit made accessible the ultraviolet part of the spectrum. In particular, the hydrogen molecule could finally be observed directly in interstellar clouds through its electronic transitions in absorption toward hot stars, first through short-duration rocket-borne instruments \citep{1970ApJ...161L..81C,1973ApJ...179L..11S} and then in extensive surveys of H$_2$ and HD with the Princeton {\it Copernicus} spectrophotometer on board the NASA OAO-2 satellite \citep{1973ApJ...181L.116S,1974ApJS...28..373S}.
The detailed study of spontaneous radiative dissociation of H$_2$ \citep{1972JQSRT..12..569S} was key to any accurate description of the atomic/molecular boundary in interstellar clouds, as first proposed by Philip Solomon and further developed by Theodore Stecher and David Williams\footnote{David Williams was one of Dalgarno's early students, and is himself a leading figure in astrochemistry} (see \S 4.2).

Non-polar interstellar molecules C$_2$H$_2$ and C$_2$ were identified through high-resolution infrared spectroscopy, the former in vibration-rotation bands and the latter in electronic transitions. 
Space probes visited several planets in our solar system and returned close-up and {\it in situ} measurements of atmospheres of planets and planetary satellites. It is important to note the NASA Atmospheric Explorer satellite program, which provided rich sets of data on the upper atmosphere of Earth \citep{1973RaSc....8..263D}. This program supported laboratory studies and theoretical research groups. Dalgarno led one such group at the Harvard-Smithsonian Center for Astrophysics, concerned with the development of models to interpret the measurements. This resulted in cross-fertilization of ideas about ion chemistry, the role of metastable atoms, and energy-loss and -deposition in dilute gases: atmospheric chemistry and astrochemistry both benefited.  

\subsection{1980s}
The identification of an exotic cyclic molecule, cyclopropenylidene (C$_3$H$_2$), in space was a triumph for astronomical spectroscopy and an incentive for further laboratory studies \citep{1985ApJ...299L..63T}. This work was carried out by the group of Patrick Thaddeus shortly before their move to the Division of Engineering and Applied Sciences and the Center for Astrophysics at Harvard in 1986, which strengthened a long tradition of laboratory astrophysics there. Improvements in high-resolution infrared spectroscopy enlarged the list of non-polar molecules in space with C$_3$ and C$_5$ (1988-1989) and CO$_2$ (1989). 

This decade saw increased attention to solid matter (dust particles) and very large molecules in space, driven in part by the all-sky photometric survey in the mid- and far-infrared carried out with the Infrared Astronomical Satellite (IRAS), a joint project of the US, the UK, and The Netherlands. Laboratory studies were beginning to address not only the optical properties of dust particles but also the chemical processes that take place in irradiated ices on their cold ($\sim 10$ K) surfaces. IRAS revealed a population of ultra-luminous galaxies powered by bursts of star formation. These were quickly understood to be extreme forms of photodissociation regions (PDR).

Several strong but unidentified infrared emission bands were seen as possible evidence  of such species as polycyclic aromatic hydrocarbons (PAH) and fullerenes. Dalgarno and co-workers investigated the role of large molecules in the ionization balance and ion chemistry in interstellar gas \citep{1988ApJ...324..553L,1988ApJ...329..418L}.

\subsection{1990s}
Arguably the most important interstellar ion, H$_3^+$, was identified spectroscopically in the ionosphere of Jupiter \citep{1989Natur.340..539D} and subsequently in interstellar gas by  \cite{1996Natur.384..334G}. Dalgarno and collaborators \citep{1973ApJ...183L.131D} had previously predicted the radio spectrum of the deuterated isotopologue H$_2$D$^+$ and discussed its observability in interstellar clouds. Following numerous attempts and tentative results, its detection finally became secure\citep{1999ApJ...521L..67S}. These two ions play leading roles in temperature-sensitive deuterium fractionation and as sensors of ionization in the overwhelmingly neutral interstellar medium. 

The Submillimeter Wave Astronomy Satellite (SWAS) and the Odin satellite took submm-wave spectroscopy into orbit outside Earth's opaque atmosphere. These made possible more sensitive searches for the elusive oxygen molecule O$_2$ and thermal, non-maser water emission, both of which are otherwise inaccessible to ground-based telescopes. SWAS enabled the identification of HF and confirmed that this rare molecule is a useful proxy for H$_2$ \citep{1997ApJ...488L.141N}. 

The Infrared Space Observatory (ISO), a satellite launched in 1995 by the European Space Agency (ESA), added considerably to our knowledge of interstellar ices and more generally to the body of infrared measurements of star-forming regions. 

\subsection{2000s}
%Dalgarno and McCray 
\cite{1973ApJ...181...95D} and \cite{1981Natur.289..656H} discussed the possible role of negative ions in interstellar chemistry long before the first anion, C$_6$H$^-$, was finally identified  spectroscopically in interstellar clouds and circumstellar envelopes \citep{2006ApJ...652L.141M}. \cite{2000IAUS..197....1D} had already anticipated that:  ``For carbon chains a substantial fraction may exist as anions as well as in neutral form.''

\subsection{2010s}
The hydroxyl ion OH$^+$ was identified in the interstellar medium \citep{2010A&A...518A..26W} via ground-based submm-wave observations\footnote{This was followed immediately by a detection of blue-violet electronic transitions in interstellar absorption \citep{2010ApJ...719L..20K}. Forty years earlier the discovery of interstellar CO at mm-wavelengths\citep{1970ApJ...161L..43W} was reported only months before interstellar CO was identified in ultraviolet absorption spectra from a rocket-borne instrument \citep{1971ApJ...164L..43S}.} shortly before the ESA Herschel Space Observatory opened up a flood of results in this waveband, including extensive observations of OH$^+$, H$_2$O$^+$, HCl$^+$, and H$_2$Cl$^+$. These ions provided important empirical tests of ion chemical schemes proposed in the 1970s. By the end of the decade, both anticipated (HeH$^+$) and unexpected (ArH$^+$) molecular ions were detected in space and the elusive O$_2$ was confirmed in interstellar clouds at a low level of abundance. 
% The CO discovery story, in more detail, is this: Wilson et al. was submitted to ApJL 1970-6-5 and published in the 1970-7 issue(!). Smith and Stecher submitted their paper 1970-12-31 slightly less than 7 months later and it was published 1971-3-1. Note that Smith and Stecher presented the first 12CO/13CO abundance ratio. Smith & Stecher appeared in vol 164 while the first mm-wave observation of isotopologues by Penzias, Jefferts, Wilson (1971) appeared later in vol 165.

\subsection{2020s}
Dalgarno and others proposed that molecular ions like CH$_3^+$ are cornerstones of interstellar ion chemistry already in the 1970s. It took more than 50 years for the existence of CH$_3^+$ to be confirmed directly by observation \citep{2023Natur.621...56B,2025A&A...696A..99Z}. The identification of the methyl cation in space was made possible by the sensitivity and high angular resolution of the spectroscopic instruments on the James Webb Space Telescope (JWST). There are several other recent revelations from JWST that attest to the legacy of Alex Dalgarno regarding the response of H$_2$ molecules to ultraviolet pumping and to interactions with cosmic rays. Strong UV illumination in a PDR appears sufficient to produce detectable H$_3^+$ without cosmic rays \citep{2025A&A...699L..13S,2025arXiv250605189G}. The infrared signature of cosmic-ray-excited H$_2$ in a dark interstellar cloud has been observed directly for the first time \citep{2025arXiv250820168B}. 

It can be argued that the chemical evolution of rotating disks of gas and dust in star- and planet-forming regions has attained the status of a paradigm in astrochemistry: it can be tested both where stars are currently forming and with reference to measurements of the planets and smaller bodies of our solar system. For example, complementary observations with the Atacama Large Millimetre/submillimetre Array (ALMA) and JWST revealed both crystalline and amorphous silicate minerals together with gaseous silicon monoxide in a system that is currently assembling solid bodies \citep{2025Natur.643..649M}.

%\subsection{}
\section{Scientific themes}
In this section, I highlight a few topics in astrochemistry to which Alex Dalgarno contributed directly.

\subsection{Chemistry and physics of the early universe}
According to the standard cosmological model, the cosmic expansion began in a Big Bang approximately $13.7\times 10^9$ years ago. Within a few minutes, the composition of normal matter was set: nuclei of hydrogen, deuterium, helium, and a small amount of lithium, plus free electrons. Approximately 380\,000 years after that event, the temperature and density had dropped sufficiently that recombination processes raised the neutral fraction to $\sim 1/2$ and the universe then became transparent to its own radiation. The average rate of recombination failed to keep pace with expansion, leaving a small fraction $\sim 10^{-4}$ of unrecombined protons and electrons as a well as a comparable fraction of molecular hydrogen and even smaller abundances of molecules like HD, HeH$^+$, and LiH.
%Despite remaining remaining mysteries about the interpretation of dark matter and dark energy, several 
%crucial parameters in the standard model are very well constrained by observation, including the
% baryonic mass fraction, the rate of expansion after the first few minutes, and the temperature of the 
%cosmic background radiation. 
As pointed out by \cite{2013ARA&A..51..163G} the framework of the standard cosmological model is so well constrained that our understanding of the chemical evolution of matter prior to galaxy formation is limited mainly by uncertainty in reaction rates. Dalgarno and collaborators investigated many of the crucial processes and their rates 
\citep{1996ApJ...458..397D,2006ApJ...646L..91D,2002JPhB...35R..57L,1993ApJ...414..672S,1996ApJ...458..401S,1998ApJ...509....1S,1998ApJ...508..151Z,2011ApJ...737...44G} 
and they helped refine models of cosmic chemistry previously introduced by %Lepp \&\ Shull 
\cite{1984ApJ...280..465L}. It is important to recognize that the presence of molecules like H$_2$ endows a dilute gas of hydrogen and helium with an ability to cool to temperatures below $10^4$ K, which would not be possible in purely monatomic gas. Low-temperature cooling is needed for the gravitational collapse and fragmentation of structures as small as stars in the early universe \citep{1967Natur.216..976S}. The most efficient formation of H$_2$ at recombination occurs via radiative attachment to form H$^-$ 
$$ {\rm H} + e^- \to {\rm H}^- + \gamma $$
followed by associative detachment from the anion
$$ {\rm H}^- + {\rm H} \to {\rm H}_2 + e^- $$
as first suggested by Dalgarno in a private communication to 
%McDowell 
\cite{1961Obs....81..240M}, who discussed the process with reference to the formation of stars in the interstellar medium (cf. also \cite{1969JPhB....2..885B}). Indeed Dalgarno himself wrote ``My involvement with H$^-$ led me into astrophysics'' \citep{2008ARA&A..46....1D}. He would return to this anion repeatedly, as late as 2010 to consider resonance-enhanced photodestruction and its implications in cosmology \citep{2010ApJ...709L.168M,2012JPhCS.388b2034M}.

It was already mentioned (\S 3.1) that Dalgarno's early quantum mechanical treatment of spin-changing collisions in hydrogen was important for understanding the excitation of the \ion{H}{I} 21 cm line. When the possibility of mapping fluctuations in the distribution of hydrogen in the early universe became recognized as one of the chief drivers of the next generation of radio telescopes, he revisited the atomic physics of spin-changing collisions \citep{2003PhRvA..67d2715Z}. See also \cite{2007MNRAS.375.1241H} and \cite{2005ApJ...622.1356Z} for discussion of some further implications.

%\subsection{Chemistry and physics of interstellar clouds, models, and photodissociation regions}
\subsection{The hydrogen molecule}
As outlined in \S 3 Dalgarno became increasingly occupied with astrophysical problems during the 1970s when many new interstellar molecules were discovered in rapid succession. In order to understand the ubiquity and relatively high abundances of these molecules, Dalgarno and others investigated networks of ion-neutral reactions, which could drive an active chemistry even at very low temperatures (10 to 100 K) rapidly enough to approach a chemical steady state within the estimated lifetimes of gas clouds at near-vacuum densities ($10^2$ to $10^4$ molecules cm$^{-3}$). His work on processes involving the hydrogen molecule was critical for the quantitative understanding of molecules in space. 

Dalgarno and collaborators performed a number of important theoretical studies of 
%the properties of and processes involving 
hydrogen molecules, beginning with Raman and Rayleigh scattering by H$_2$ \citep{1962MNRAS.124..313D,1962ApJ...136..690D}. Rayleigh scattering is currently of interest in relation to atmospheres of exoplanets. The influential article on the theory of scattering by a rigid rotator \citep{1960RSPSA.256..540A} led to the early distorted-wave and close-coupling calculations of the rotational excitation of H$_2$ by collisions with electrons, \citep{1965PPS....85..679D}, H \citep{1966PPS....88..611D}, He, and H$_2$ \citep{1967PPS....90..609A}. Much later this would be extended further to include reactive collisions between H and H$_2$ \citep{1994ApJ...427.1053S}, to include rotation-vibration coupling \citep{1999ApJ...514..520B}, and to examine effects of improved interaction potentials \citep{1997ApJ...489.1000F}. 

Quantitative spectroscopy in astrophysics requires accurate transition probabilities. \cite{1968ApJ...154L..95D} computed band oscillator strengths for the Lyman system in the ultraviolet spectrum of H$_2$ and extended this to include the Werner system and the deuterated isotopologues \citep{1969AD......1..289A}. Thus the necessary data were already available for the interpretation of the first direct observations of interstellar H$_2$ beginning in 1970 (see \S 3.2). 

Although the existence of interstellar H$_2$ had been predicted \citep{1966ARA&A...4..207F}, there remained some uncertainty about the rates of formation and destruction of the molecule. Because the direct radiative association of ground-state H atoms to form H$_2$ is a strongly forbidden process and because formation via H$^-$ is slow (\S 4.1), it was expected that association of H atoms on the surfaces of dust particles must be the dominant source of interstellar H$_2$. It had been pointed out by Philip Solomon\footnote{cited as a private communication 1965 by Field, Somerville, \&\ Dressler (1966)} that the fastest destruction process  of interstellar H$_2$ was likely to be spontaneous radiative dissociation into the continua of the B $^1\Sigma^+_{\rm u}$ and C $^1\Pi_{\rm u}$ states following absorption of UV starlight in lines of the B$\gets$X (Lyman) and C$\gets$X (Werner) systems. The efficiency of this process was uncertain until \cite{1970ApJ...160L.107D} and \cite{1972JQSRT..12..569S} did accurate computations of the transition probabilities to both bound and unbound states. Related work also solved the mystery of the new continuous emission spectrum of H$_2$ discovered by Herzberg \citep{1970ApJ...162L..49D}. \cite{1973ApJ...186..165S} went on to compute the kinetic energy of dissociated atom pairs in this process, which is a heating mechanism in interstellar clouds. 

Early on Dalgarno appreciated that accurate probabilities of quadrupole vibration-rotation transitions were needed for analyzing the excitation of interstellar H$_2$ and that these might someday be observable \citep{1969JAtS...26..946D,1972ApJ...174L..49D,1977ApJS...35..281T,1998ApJS..115..293W}. 
These transition probabilities have been in routine use to analyze infrared spectra.

\subsection{Thermal balance: heating and cooling processes}
Some of Dalgarno's earliest forays into astrophysics concerned cooling mechanisms in interstellar gas. Aside from adiabatic expansion, the most important cooling mechanisms are microscopic processes that remove kinetic energy from the gas and convert it into radiation that escapes. Especially important are inelastic collision processes that excite low-lying fine-structure levels in atomic ions and excited rotational states in common molecules \citep{1964ApJ...140..800D,1971ApJ...168..161W,1972ARA&A..10..375D,1975RSPSA.342..191C,1975JChPh..62.4009C,1975ApJ...200..419O}. Decades later Dalgarno and collaborators re-visited several of these processes through use of improved theoretical methods and interaction potentials \citep{2005ApJ...620..537B,2006ApJ...647.1531K,2007ApJ...654.1171A,2007A&A...475L..15S}.
Subsequent improvements in observations at mm and far-infrared wavelengths confirmed that such transitions cause some of the most intense emission lines seen from interstellar matter.
A related consideration is the collisional excitation of low-lying rotational states in molecules like CN, the excitation of which would otherwise be locked by radiative coupling to the cosmic microwave background (CMB) radiation \citep{1969A&A.....2..451C,1971A&A....13..331A,1974PhRvA..10..788C}. In addition, sequences of exoergic ion-neutral reactions can add net kinetic energy to the gas and constitute a form of chemical heating \cite{1974ApJ...192..597D}.

\subsection{Interstellar ion chemistry}
The flood of discoveries of interstellar molecules in the 1970s (\S3.2) stimulated many proposals of networks of 
chemical reactions to explain their abundances. Early on it was recognized that exoergic ion-neutral reactions 
might play a key role because they remain rapid at low interstellar temperatures  \citep{1951ApJ...113..441B,1972ApJ...178..389S,1973ApJ...185..505H,1973ApJ...183L..17W,1974ApJ...188...35W}. \cite{1973ApL....15...79B} suggested that a radiative association process
$$ {\rm C}^+ + {\rm H}_2 \to {\rm CH}_2^+ + \gamma $$
could be rapid enough to explain the abundance of interstellar CH and to initiate the formation of many other carbon-containing molecules, including widespread formaldehyde \citep{1973NPhS..245..100D}. The connections 
among cosmic-ray ionization, deuterium isotope-exchange reactions, and molecular abundances and excitation 
were explored \citep{1973ApJ...184L.101B,1973ApL....14...77D,1976RPPh...39..573D}.

Despite some early successes of ion-neutral reaction schemes, a long-standing mystery surrounded the abundance of interstellar CH$^+$. Methylidyne CH and its cation CH$^+$ were the first two molecules to be identified in interstellar space  \citep{1937PASP...49...26D,1941ApJ....94..381D}. The derived interstellar abundances of these two species were found to be comparable, even though the ion should be destroyed more rapidly by exoergic reactions with H$_2$ and H than the neutral is destroyed by reactions with ions or by photodissociation. Attempts to explain CH$^+$ with modifications to the simple ion chemistry \citep{1975ApJ...199..633B} gave way to enthusiasm for high-temperature chemistry in components of gas heated by shock waves \citep{1986MNRAS.220..801P}. My view at the present time is that turbulent dissipation regions (TDR) of the dilute molecular gas or
multi-phase turbulent diffusion successfully explain CH$^+$ formation via the 
otherwise endoergic reaction 
$$ {\rm C}^+ + {\rm H}_2 \to {{\rm CH}}^+ + {\rm H} - \{ 0.43 {\rm eV}\} \;\;\; $$
\citep{2009A&A...495..847G,2023A&A...669A..74G}.
This is supported by independent observations of the distribution functions of velocities in nearby molecular clouds, which are in harmony with intermittent turbulent motions 
following injection of the appropriate total power on the largest scale. It is interesting that pure rotational emission lines of CH$^+$ can now be observed in the distant universe \citep{2017Natur.548..430F,2021MNRAS.506.2551V}. In strongly illuminated photodissociation regions (see the following subsection) the presence of ultraviolet-excited H$_2(v\geq 1)$ provides an alternative to high kinetic temperature as the explanation for rapid formation of CH$^+$, as shown by very recent analysis of JWST observations  \citep{2025A&A...696A..99Z}. 

\cite{1973ApJ...183L..21D} and \cite{1977ApJ...212..683O} noted some examples of associative ionization reactions, which might be important even at low temperatures in interstellar gas. This allows a type of chemical 
ionization process that begins with neutral reactants and produces an ion and a free electron, for example
$$ {\rm CH} + {\rm O} \to {\rm HCO}^+ + e^- \;\;\;. $$
The context is interesting. \cite{1973ApJ...183L..21D} estimated the abundance of the formyl ion HCO$^+$ 
that could be formed this
way in interstellar clouds at around the same time that  \cite{1973ApJ...185..505H} 
completed their seminal study of interstellar ion chemistry. Interest in HCO$^+$ was high at this time because Klemperer had earlier suggested it as the
carrier (``X-ogen'') of a strong, unidentified mm-wave emission line near a frequency of 89.2 GHz 
\citep{1970Natur.227.1230K,1974ApJ...188..255H}.  
The definitive laboratory spectroscopy needed to confirm the HCO$^+$ identification came a few years later \citep{1975PhRvL..35.1269W}. Even if the associative ionization process is not the main source of interstellar
HCO$^+$, such reactions merit further attention. 

As mentioned in \S4.1, Dalgarno and collaborators recognized the role of H$^-$ in the chemistry of the early universe and \cite{1973ApJ...181...95D} were among the first to propose that anions might participate in interstellar chemistry more generally. At the time of this writing, eight species of carbon-chain anions have been identified spectroscopically in interstellar clouds or circumstellar envelopes. The abundances of these anions and attempts to interpret them have been summarized most recently by \cite{2023A&A...677A.106A}.

Dalgarno and collaborators were among the first to investigate possible interstellar chemistries of rarer elements:
silicon \citep{1977ApJ...213..386T}, sulfur \citep{1974ApJ...187..231O}, chlorine \citep{1974ApJ...192L..37D}, and titanium \citep{1977ApJ...212..683O}.

\subsection{Models of clouds and photodissociation regions (PDR)}
Progress was extremely rapid in the 1970s. Within a few years of the first mm-wave observations of CO in interstellar space (1970), this molecule was known to be abundant and widespread, and was already being used as a proxy for the distribution of molecular hydrogen. In order to quantify the cooling effect of CO in relation to the gravitational collapse of gas clouds to form stars, it was necessary to design models that incorporated the best available atomic and molecular data. Among the first of these was \cite{1975ApJ...199...69D}, along with \cite{1973ApJ...179L.147G} and \cite{1974ApJ...189..441G}.

An ideal molecular cloud in our Milky Way Galaxy has a sharp boundary between mostly atomic and mostly molecular composition because the most abundant molecule, H$_2$, is destroyed mainly by spontaneous radiative dissociation following absorption of UV starlight in sharp lines (see above \S 4.2 and \cite{1967ApJ...149L..29S}). Under typical interstellar conditions this absorption from the most populous rotational levels of 
%the ground vibronic state of 
H$_2$ begins to saturate when the molecular column density exceeds $N({\rm H}_2)\approx 10^{13}$ cm$^{-2}$. Upon saturation, the absorption rate decreases exponentially with increasing depth into the cloud. Thus the photo-destruction rate of H$_2$ diminishes from its unshielded value of approximately $5\times 10^{-11}$ s$^{-1}$ in the average background starlight (lifetime $\sim 600$ years) to values below $10^{-16}$ s$^{-1}$, at which point the destruction by pervasive cosmic rays takes over.  This effect of self-shielding by H$_2$ takes place within a boundary zone that is a small fraction of the full size of a typical molecular cloud. The H/H$_2$ boundary layer is thus regulated by ultraviolet starlight in the wavelength interval 92 to 110 nm, which coincidentally includes the near-threshold photoionization continuum of atomic carbon, the predissociated bands of carbon monoxide, and all but one of the Lyman series lines of H. In the neutral interstellar medium outside the ionized nebulae around hot stars, H-ionizing photons (wavelengths shorter than 92 nm) have already been excluded. This photon-limited boundary layer in H/H$_2$  and C$^+$/C/CO is also a zone in which the  gas is predominantly heated by photoelectric ejection from dust particles. The atomic/molecular boundary layer is now commonly known as a photodissociation region (PDR), a term and abbreviation coined by \cite{1985ApJ...291..722T}. Dalgarno himself preferred the label `photon-dominated region' to acknowledge the myriad other processes that make PDR stand out against their darker backgrounds. 

The observations of interstellar atoms and molecules in the 1970s already signaled a demand for integrated theoretical models to explain:
\begin{itemize}
 \item observed column densities of H$_2$ and H
 \item rotational excitation of H$_2$, with lowest levels at Boltzmann excitation temperatures $\sim 70$ K and $J=2-8$ at 200-300 K 
 \item relative abundances of other small molecules such as HD, CH, CH$^+$, CO, etc.
 \item thermal balance, heating and cooling rates
 \item degree of ionization.
\end{itemize}
Dalgarno recognized that the balance between H and H$_2$ must be closely related to the processes that excite H$_2$ in diffuse molecular clouds, because on average every UV-induced dissociation event is accompanied by fluorescence to bound, excited vibration-rotation levels of the ground state. Bound-bound fluorescence dominates (ca. $85\%$ in the spectrum of background starlight) and the ensuing radiative cascade via quadrupole vibrational-rotational transitions produces a characteristic emission spectrum. This infrared response of H$_2$ to UV starlight was explored in detail by \cite{1976ApJ...203..132B} and incorporated into physical and chemical models of diffuse molecular clouds \citep{1977ApJS...34..405B,1978ApJ...224..448B}. Similar models were developed by other groups \citep{1974ApJ...193...73G} with improvements in the treatment of the self-shielding of H$_2$ \citep{1979ApJ...227..466F}.
In the following decade, Dalgarno's former students developed comprehensive models of diffuse and translucent molecular clouds \citep{1986ApJS...62..109V,1989ApJ...340..273V}, which incorporated a detailed treatment of the dissociation and chemistry of CO \citep{1988ApJ...334..771V} in order to keep up with the growing body of observational data. It is important to emphasize that diffuse and translucent molecular clouds illuminated by background
ultraviolet starlight are low-flux photodissociation regions that afford an even wider variety of observational tests 
than their counterparts in complex, high-flux star-forming regions.  \cite{1987ApJ...322..412B} examined the spectrum of radiatively excited H$_2$ in PDR in more detail, and \cite{1989ApJ...338..197S} investigated how the UV-excited infrared emission of H$_2$ is altered at high densities when inelastic collisions partly interrupt the quadrupole radiative cascade and went on to consider chemical diagnostics of dense PDR \citep{1995ApJS...99..565S}. 

Meanwhile Dalgarno and collaborators investigated photodissociation and photoionization processes in other crucial molecules, including H$_2$ \citep{1975ApJ...195..819F,1975ApJ...200..788F}, CH$^+$ \citep{1978CP.....32..301U,1979CPL....63...22U}, NaH and LiH \citep{1978ApJ...224..444K}, HCl \citep{1982JChPh..77.3693V},  HCl$^+$ \citep{1991JChPh..95.9009P}, OH \citep{1983JChPh..78.4552V,1983JChPh..79..873V,1984ApJ...277..576V,1984JChPh..81.5709V}, etc. The investigation of effects of scattering and absorption by dust on radiative transfer models provided valuable results on the depth-dependence of photo-processes through molecular clouds \citep{1981ApJ...243..817R,1991ApJS...77..287R}. Scaling relations based on those methods continue to be applied and can be found in the Leiden Database \citep{2017A&A...602A.105H}.\footnote{The {\it Leiden database of photodissociation and photoionization of astrophysically relevant molecules} can be found at \href{https://home.strw.leidenuniv.nl/~ewine/photo/}{{\tt https://home.strw.leidenuniv.nl/{$\sim$}ewine/photo/}}.}
 
The concept of a molecular photodissociation region can be generalized to include an X-ray-dominated input spectrum. A non-thermal X-ray spectrum broadens the atomic / molecular transition zone and enables it to harbor a wider variety of ions. Among the earliest explorations of the structure of an 
X-ray-dominated region (XDR) were those of \cite{1995ApJ...446..852G,1996A&A...306L..21L,1997IAUS..178..141S,1997ApJ...481..282T} and \cite{1996ApJ...466..561M}.
Later developments in PDR studies include the publication of several public computer codes. These were subjected to a rigorous comparative benchmarking exercise \citep{2007A&A...467..187R}. 
Worthy of note is the {\it Meudon PDR code} \citep{2006ApJS..164..506L}, which incorporates a vast array of molecular processes and a rigorous treatment of the abundance and excitation of H$_2$, HD, C, C$^+$, and CO. Most recently this code has been used to interpret sensitive infrared spectra of the Orion Bar PDR at high spatial resolution,\footnote{An early-release science project of JWST called {\it PDRs4All} has already produced sixteen published papers that describe detailed tests of the most sophisticated models of PDR\citep{2022PASP..134e4301B}.
The project and its products are described at \href{https://www.pdrs4all.org/}{{\tt https://www.pdrs4all.org/}} .}
 which shows --- among other things --- how vibrationally excited H$_2$ is involved in the formation of other observed molecules. The {\it Cloudy code} was originally developed Gary Ferland and collaborators to model astrophysical plasmas, but now includes extensive treatments of molecular processes in PDR and other weakly ionized plasmas \citep{2005ApJS..161...65A}. 
It is fair to assert that PDR are the brightest manifestations of molecular material in the universe. Nearby examples within our own Galaxy highlight the places where recently formed, massive stars continue to interact with their natal clouds. In the mid-1980s, far-infrared surveys of the sky revealed a class of so-called {\it ultra-luminous infrared galaxies} (ULIRG) which radiate total powers comparable to those of quasi-stellar objects (QSO or quasars). The ULIRG phenomenon is almost always associated with close collisions and mergers of gas-rich galaxies. Mergers of gas-rich systems can concentrate a large mass ($> 10^{10}$ M$_{\odot}$) of molecular gas in a relatively small volume faster than would be possible through gravitational friction alone, precisely because the dissipative component --- that is, the molecular gas and dust --- is so effective in turning kinetic energy of gas motions into radiation. The high luminosity of an ULIRG is typically generated in a burst of star-formation (as much as 100 solar masses per year) within a relatively small volume. The PDR of the star-forming gas re-processes a large fraction of the total radiant energy into the infrared waveband; as a result, such luminous star-forming galaxies are among the most distant sources currently detectable in astronomy. As emphasized 
by  \cite{2022ARA&A..60..247W} in a recent review, ``PDR and XDR tracers are now routinely detected on galactic scales over cosmic time. This makes it possible to link the star-formation history of the Universe to the evolution of the physical and chemical properties of the gas.'' As detailed above, we can hope to understand these phenomena in detail in large part following the early, accurate work of Dalgarno and collaborators on the crucial microscopic processes that control PDR.

\subsection{Cosmic rays and interstellar chemistry}
% Prasad and Tarafdar
\cite{1983ApJ...267..603P} noted that secondary electrons produced via cosmic ray ionization in an interstellar cloud will excite H and H$_2$ to radiate internal ultraviolet radiation, which will further photodissociate and photoionize other species. This is  important in the gas-phase chemistry in a dense molecular cloud and is now routinely incorporated into chemical models \citep{1987ApJ...321..383L,1987ApJ...323L.137G}. Dalgarno and collaborators performed detailed calculations of the efficiencies in the Prasad-Tarafdar mechanism \citep{1987ApJ...320..676S,1989ApJ...347..289G}. 
These efficiencies have been improved and updated most recently by \cite{2024A&A...682A.131P}.

\section{Planetary atmospheres, aeronomy, and comets}
Some of Dalgarno's early insights into interstellar chemistry were informed by his experience in studying the atmospheres of Earth and other planets. The dilute gases of the upper atmosphere and interstellar space have similarities even though the typical pressure and elemental compositions are quite different. Photoionization and photodissociation of abundant molecules in the upper atmosphere produce ions and radicals that drive a rapid chemistry. These processes also deposit energy and yield excited products, with resulting detectable 
airglow emissions.

Beginning in the 1970s space probes made possible the {\it in situ} and close-up measurement of the outer 
atmospheres of planets and planetary satellites. More recently, following the discovery 
of planets orbiting other normal stars in 1995, it has become possible to observe atmospheres of exoplanets. 
Dalgarno and collaborators (notably Jane Fox) 
established a basis for understanding the processes that govern the chemistry
and physics of the upper atmospheres of planets. In particular, that work concerned atmospheric heating
and ionization \citep{1979P&SS...27..491F,1977P&SS...25...71F,1979JGR....84.7315F,1981JGR....86..629F}, airglow \citep{1973P&SS...21..383Z,1977JGR....82.1615F,1996ApJ...462..502L}, and escape processes \citep{1980P&SS...28...41F,1983JGR....88.9027F}.  

Investigations of photodissociation of OH with Ewine van Dishoeck led to accurate calculations of the lifetime of this important radical and its deuterated form in comets \citep{1983Icar...56..184S,1984Icar...59..305V}. 
Following the surprising discovery of extreme ultraviolet (EUV) and X-ray emission from Comet C/Hyakutake 1996 B2 \citep{1996Sci...274..205L}, 
Dalgarno's former student Thomas Cravens identified its source as highly stripped heavy ions of the solar wind undergoing charge exchange with neutral species in the cometary coma, the highly excited products of which readily emit EUV and X-ray photons \citep{1997GeoRL..24..105C,2000AdSpR..26.1443C}.  \cite{2000JGR...10518351K} investigated the ion-charge-transfer spectrum in detail. A series of further papers  followed \citep{2001ApJ...554L..99K,2002Icar..160..437K,2002PhRvA..66f4701R,2003ApJ...585L..73K,2004ApJ...617.1347P,2006ApJ...650..461S}
and showed how X-ray emission of comets could be used as a diagnostic of solar-wind composition and interaction with dilute atmospheres. A related study of X-ray emission of the dark Moon caused by solar-wind charge transfer \citep{2004ApJ...607..596W} suggested that such processes in the heliosphere contribute significant contamination of the diffuse background in X-ray astronomy \citep{2006A&A...460..289K,2009SSRv..143..217K}
\footnote{Although this work is not strictly speaking ``astrochemistry'', it does illustrate the irony of serendipity in the history of astronomy. X-ray astronomy as we know it began by accident during a rocket experiment in 1962 to detect fluorescence from solar X-rays striking the Moon. The detector swept across the Moon detecting nothing but found an astonishingly bright point-source in the constellation Scorpius, called Sco X-1 ever since  \citep{1962PhRvL...9..439G,1974ASSL...43....1G}. This opened a new window on the universe and eventually revealed high-energy phenomena associated with neutron stars, black holes, and hot plasma in clusters of galaxies. Riccardo Giacconi shared the Nobel Prize in Physics in 2002 for the development of X-ray astronomy. It took many decades after 1962 to achieve sufficient sensitivity to detect X-rays from the original target, the Moon. }.
Similar processes have implications for X-ray aurorae in Jovian planet atmospheres \citep{2006GeoRL..3311105K,2008JGRA..113.8229K,2010JGRA..115.7102H}.

\section{Astrophysical shock waves and supernovae}
Although much of the molecular component of interstellar and circumstellar matter is cold (kinetic temperature $<100$ K), stellar winds and explosions create excess pressure and drive shock waves through the gas. Even in undisturbed regions, there are motions that can be described as turbulent, over a range of scales from the microscopic dissipation length up to the size of a giant molecular cloud ($\sim 10^{17}$ m). The motions can be perceived though Doppler shifts and kinematical spectral line broadening. Dalgarno and collaborators did some important work on the structure and impacts of molecular shocks. \cite{1979ApJ...233L..25D,1982ApJ...255..176R} showed that molecules can survive destruction in faster shocks than previously thought because at low interstellar densities the rates of collisional dissociation depart from the values that would be measured in conditions near thermodynamical equilibrium. Shock waves raise the temperature of gas and thus open reaction channels that would be endothermic in colder, quiescent gas. Consequently there are some molecular abundances that might provide signatures of shock-induced chemistry \citep{1980ApJ...236..182H,1981RSPTA.303..513D}.
Interstellar conditions also permit C-type (continuous) magnetic shock waves in which magnetic field strengths as low as $\sim 1\; \mu$G induce drag between the ions and neutrals. \cite{1983ApJ...264..485D} were among the first to explore the structure and effects of such shocks in molecular clouds, with later follow-up \citep{1986MNRAS.220..801P,1987ApJ...317..432G,1988MNRAS.235..621P,1989ApJ...340..869N,1989ApJ...344..251N}.

Supernova explosions deposit enormous amounts of radiant and kinetic energy into interstellar space. It is also thought that expanding supernova remnants are the sources of cosmic rays. The large-scale properties of the Galactic interstellar medium have long been thought to be regulated at least in part by sporadic supernova events: at an average rate of the order of one per century, a typical location in the Galactic disk would be exposed to such an outburst every 10 million years. Early efforts to describe the global heating and ionization of the interstellar medium \citep{1969ApJ...155L.149F,1969ApJ...158..173G}
identified two thermally stable phases. As a partial alternative to this steady-state view 
 \cite{1970ApL.....6..237B} and \cite{1972ApJ...174..365J} proposed a time-dependent model of the interstellar medium. Such global models evolved further in subsequent decades \citep{1977ApJ...218..148M,2005ARA&A..43..337C}. 

The supernovae themselves are important factories of chemical elements through rapid nucleosynthesis.
They may be important sources of molecules and dust in the early universe \citep{2008ApJ...683L.123C}.  
Following the outburst of SN 1987A in the Large Magellanic Cloud --- the nearest and best studied supernova event in modern time --- it became clear that supernova atmospheres contain molecules and may contribute to the rapid condensation of some dust particles. Among the first to analyze the chemistry in the ejecta of supernovae were \cite{1990ApJ...349..675D,1990ApJ...358..262L,1992ApJ...396..679L,1994ApJ...428..769L,1995ApJ...454..472L,1996ApJ...471..480L,1997Ap&SS.251..375D,1998ApJ...500.1049Y,1999Sci...283.1290C}. Subsequent work proposed a tentative identification of emission due to H$_3^+$ in SN 1987A \citep{1992Natur.355..420M} and explored further aspects of ion chemistry in such an environment \citep{1998ApJ...500.1049Y}.
Even 30 years after the explosion was first observed, the chemistry of SN 1987A was still being explored \citep{2017MNRAS.469.3347M,2017ApJ...842L..24A}. For example, evidence has been presented that H$_2$ in the ejecta of SN 1987A must be formed in the gas phase rather than on dust particles \citep{2019ApJ...873...15L}.

\section{Institutions and human networks}
In the preceding sections I have tried to show how much of the early development of astrochemistry flowed directly
from the scientific research of Alexander Dalgarno. His influence on the subject  extended far beyond that, especially through his many efforts to initiate and to strengthen institutions and organizations. Within a few years of 
joining the faculty at Harvard University, Alex took up leadership positions in the Department of Astronomy and the Harvard College Observatory upon the departure of Leo Goldberg in 1971. He played an important part in the
consolidation of the Harvard College Observatory and the Smithsonian Astrophysical Observatory into the
Harvard-Smithsonian Center for Astrophysics (CfA), which emerged in 1973 under a single director, George Field,
as one of the largest astronomical 
research organizations in the world. 
Dalgarno served as Editor of The Astrophysical Journal Letters from 1973 through 2002 (see \S 3). 
Keenly aware that atomic physics was underfunded 
and fragmented in the USA, Dalgarno persuaded the National Science Foundation to support an Institute for Theoretical Atomic, Molecular, and Optical Physics (ITAMP), to be hosted by the Harvard Department of Physics and CfA. Since its founding in 1988, ITAMP has remained a valuable resource. 

In 1979 a symposium on interstellar molecules took place at Mount Tremblant, Canada, under the auspices of the International Astronomical Union (IAU). At a time of rapid discovery (cf \S 3.2), this meeting brought together not
only astronomers, but also laboratory spectroscopists and chemists. It became clear to Alex Dalgarno and others 
that the IAU with its global reach and symposium series could provide a mechanism for nurturing a community of
astrochemistry. He led the effort to establish an official IAU Working Group in astrochemistry with the idea to 
organize an international symposium every five to six years, so that progress in the field could be monitored and
encouraged. This series of meetings has been extremely important in establishing a community of researchers 
and an evolving sense of identity for the field (Table 1).
\begin{table}   % [htp]
\caption{IAU Symposia on astrochemistry}
\begin{center}
\begin{tabular}{rlll}
\hline
 IAUS & Year & Location & Title \\  %& Publication  \\
 \hline
 87 & 1979 & Mt Tremblant, Canada & {\it Interstellar Molecules} \\  %& \cite{1980IAUS...87.....A} \\
 120 & 1985 & Goa, India & {\it Astrochemistry} \\ %& \cite{1987IAUS..120.....V} \\
 150 & 1991 & Sao Paulo, Brazil & {\it Astrochemistry of cosmic phenomena} \\ %& \cite{1992IAUS..150.....S} \\
 178 & 1996 & Leiden, The Netherlands & {\it Molecules in astrophysics: probes and processes} \\ 
 % & \cite{1997IAUS..178.....V} \\
 197 & 1999 & Sogwipo, South Korea & {\it Astrochemistry: From Molecular Clouds} \\
  &&& {\it to Planetary Systems} \\
 %& \cite{2000IAUS..197.....M} \\
  231 & 2005 & Pacific Grove CA, USA & {\it Astrochemistry: Recent Successes} \\
  &&& {\it and Current Challenges} \\
  % & \cite{2005IAUS..231.....L} \\
  280 & 2011 & Toledo, Spain & {\it The Molecular Universe} \\
  %& \cite{2011IAUS..280.....C} \\
  332 & 2017 & Puerto Varas, Chile & {\it Astrochemistry VII: through the cosmos} \\
  &&& {\it from galaxies to planets} \\
  %& \cite{2018IAUS..332.....C} \\
  383 & 2023 & Traverse City MI, USA & {\it Astrochemistry VIII: From the First Galaxies} \\
  &&&  {\it to the Formation of Habitable Worlds} \\
  % & \href{https://events.mpe.mpg.de/event/14/overview}{Astrochemistry VIII} \\ 
 \hline 
\end{tabular}
\end{center}
\label{Table 1}
\end{table}
The symposium proceedings that resulted form a valuable record of the development of the subject 
\citep{1980IAUS...87.....A,1987IAUS..120.....V,1992IAUS..150.....S,1997IAUS..178.....V,2000IAUS..197.....M,2005IAUS..231.....L,2011IAUS..280.....C,2018IAUS..332.....C}. A broad geographical distribution is evident.

\section{Astrochemistry post-AD}
It is far beyond the scope of this article to review recent developments in astrochemistry, even restricted to those that have been directly influenced by the scientific work of Alex Dalgarno. It is perhaps worthwhile to 
outline some important areas of astrochemistry that lie outside his own research interests. Dalgarno largely avoided 
chemistry in the solid phase, aside from the obvious role of dust particle surfaces in catalysing the formation
of interstellar H$_2$. Already in the 1980s, J.~Mayo Greenberg and his group in Leiden performed experiments 
to simulate the conditions in interstellar space (exceedingly low pressure, cryogenic temperature, and exposure to 
radiation) under which gaseous molecules freeze out onto cold surfaces. 
They found evidence of chemical activity in
ices, including the growth of large molecules, at a time when there was growing evidence from infrared 
spectroscopy of solid state absorption features of CO, CO$_2$, H$_2$O, and other species
in the interstellar medium.
The hypothesis that large carbonaceous molecules like polycyclic aromatic hydrocarbons (PAH) are 
responsible for the once-mysterious infrared emission bands has been largely confirmed.
Eric Herbst and others incorporated solid-phase reactions into hybrid gas-grain interstellar chemical networks. 
Processes that control the nucleation and growth of solid particles in stellar winds and in the interstellar medium
have received much attention over the years. Dalgarno's pioneering work on molecules in supernova
remnants is part of the story of their role in the formation of dust.
Thanks to telescopes like ALMA, NOEMA, and JWST,  it is now possible to resolve the structures of planet-forming 
disks around young stars on a comparable scale to the orbits of the terrestrial planets of our solar system.
This allows astronomers to study the composition of gas and dust in these systems, to infer how dust 
agglomerates to form planetesimals and planets, and to compare chemical evolution of planets in general with
the 4.6-Gyr-old record of planet-formation in our own system. In order to understand the progression from an
interstellar cloud on the brink of gravitational collapse, through an accretion-fueled protostar, to the assembly of 
planets in a rotating disk of gas and dust, we must follow the complex chemistry that flows back and forth
between solid and gaseous states.  
It could be argued that the most important ``discovery'' in astrochemistry is the recognition of the chemical
evolution that regulates the physical development of stars and planets. For a series of reviews see
\cite{1993QJRAS..34..213H},
\cite{1998ARA&A..36..317V}, 
\cite{1999ARA&A..37..311E},
\cite{2000ARA&A..38..427E},
\cite{2007ARA&A..45..339B},
\cite{2009ARA&A..47..427H},
\cite{2019ARA&A..57..113A},
and \cite{2020ARA&A..58..727J}.
 
\section{Personal reflections and summary}
I entered Harvard College as a first-year undergraduate in 1967, coincidentally the same year as Alex Dalgarno joined the Harvard faculty as professor. Our paths finally crossed four years later when he agreed to supervise my PhD studies during the period 1971-1975. At that time I already identified as an astronomer and joined Dalgarno's research group from that orientation, glad of the opportunity to become part of a growing cross-disciplinary field.

Alex Dalgarno had a remarkable ability to formulate timely research projects for students and post-doctoral researchers. He mentored young scientists from a wide range of backgrounds: astronomy, physics, chemistry, atmospheric science, and applied mathematics. The ambience of his research group was naturally interdisciplinary. 
A key feature of Dalgarno's scientific genius was a highly disciplined imagination that could perceive
quantum phenomena and give them rigorous quantitative expression, predictions which could then be tested
through observation and experiment. He breathed life into astrochemistry through this
ability to find the connections between the unseen on the quantum level 
and the unseen throughout the cosmos.  He applied 
the same careful analysis through clear, succinct words both to scientific research and to the practical problems of supervising students, leading research institutions, editing journals, and promoting international organizations.

\bibliographystyle{aa}
\bibliography{references.bib}

\end{document}